\documentclass[fleqn,usenatbib]{mnras}

\usepackage{newtxtext,newtxmath}

\usepackage[T1]{fontenc}

\DeclareRobustCommand{\VAN}[3]{#2}
\let\VANthebibliography\thebibliography
\def\thebibliography{\DeclareRobustCommand{\VAN}[3]{##3}\VANthebibliography}

\usepackage{graphicx}	
\usepackage{amsmath}	

\newcommand{\likeli}[2]{\mathcal{L}(#1 |\thinspace #2)}

\title[The hot nuclear EOS from post-merger GWs]
{Sub-threshold post-merger gravitational waves can constrain the hot nuclear equation of state}

\author[F.~H.~Panther \& P.~D. Lasky]{
Fiona H. Panther,$^{1,2}$\thanks{E-mail: fiona.panther@uwa.edu.au}
Paul D. Lasky,$^{2,3}$
\\
$^{1}$Department of Physics, University of Western Australia, Crawley WA 6009, Australia\\
$^{2}$OzGrav: The ARC Centre of Excellence for Gravitational-wave Discovery\\
$^{3}$School of Physics and Astronomy, Monash University, VIC 3800
}

\date{Accepted XXX. Received YYY; in original form ZZZ}

\pubyear{2025}

\begin{document}
\label{firstpage}
\pagerange{\pageref{firstpage}--\pageref{lastpage}}
\maketitle

\begin{abstract}
We show how to coherently combine information from a population of sub-threshold, gravitational-wave binary neutron star post-merger remnants. Although no individual event in our synthetic population can be claimed as a confident detection, we show how to statistically determine the fraction of merger events that promptly collapse to form a black hole, compared to those for which a neutron star survives the merger for at least tens of milliseconds. This fraction, when combined with information about the neutron star mass distribution gleaned from the inspiral portion of the signals, provides an indirect measure of the neutron star maximum mass. Using conservative measures of the post-merger waveforms, we show that 50-70 events with binary neutron star inspiral measurements can be combined to give an $11-20\%$ fractional uncertainty on the maximum mass of rapidly rotating, hot neutron stars, which can potentially be turned into a $12-21\%$ fractional constraint on the Tolman-Oppenheimer-Volkoff mass. We discuss how this measure of the hot nuclear equation of state can be combined with information of cold neutron stars to see the effect of temperature on physics in the densest regions of the Universe by providing indirect evidence for first-order phase transitions in neutron star interiors.
\end{abstract}




\section{Introduction}
Gravitational waves from the remnants of binary neutron star mergers provide one of few places in the Universe to study the hot nuclear equation of state. Stable or pseudo-stable neutron star remnants with lifetimes longer than tens of milliseconds are expected to form from a significant fraction of these mergers~\citep{gao16,margalit19,zou25}. Such merger remnants can exceed the maximum allowed mass of a cold, non-rotating or uniformly rotating neutron star by 30-70\%~\citep{baumgarte00,Bauswein2013}. This temporary support against gravitational collapse can be lost though processes such as the quenching of differential rotation~\citep{Shapiro2000}, neutrino emission, and gravitational-wave emission~\citep{Hotokezaka2013}.

Although the instantaneous strain of the high-frequency ($\sim$kHz) gravitational-wave signals are much louder than the late inspiral of the merging binary neutron stars, the measurable signal-to-noise ratio of the post-merger signal is potentially an order of magnitude lower due to the poor sensitivity of detectors to kHz emission and the short duration of the signals. Consequently we may not conclusively detect post-merger gravitational waves until gravitational-wave observatories such as Cosmic Explorer~\citep{evans21}, Einstein Telescope~\citep{hild11}, or NEMO~\citep{ackley20} come online in the late 2030s. Even then, these observatories are predicted to observe $2-10$ post-merger signals at signal-to-noise ratio $\mathrm{SNR}>8$ per year~\citep[e.g.,][]{,martynov19,breschi22}, and claiming an unambiguous detection may be challenging due to the characteristics of gravitational-wave detector noise at those frequencies~\citep{panther23}. Following the detection of BNS mergers GW170817 and GW190425, there were numerous searches for gravitational waves emitted by the merger products \citep{GW170817postmerger, GW170817postlong, GW190425, Grace2024, Abchouyeh2023}, however these have only placed upper limits on the strain amplitude and energetics of any putative emission.

We expect many binary neutron star mergers to emit post-merger signals that are below the threshold for detection in gravitational-wave observatories. As impatient souls, we would like to gain insight into the hot nuclear equation-of-state from the collective sum of these sub-threshold observations. In this work, we develop a technique for calculating the maximum, non-rotating neutron star mass, the so-called Tolman-Oppenheimer-Volkoff (TOV) mass~\citep{Tolman39,Oppenheimer39}, using information from an ensemble of sub-threshold post-merger gravitational-wave signals. We show how to calculate the fraction of binary neutron star mergers that result in prompt collapse to a black hole, cf. some short- or long-lived neutron star that necessarily emits kHz gravitational waves. This fraction, combined with the binary neutron star mass distribution as determined by the inspiral component of the signals, provides a measurement of the maximum \textit{rotating} mass threshold for hot neutron stars. This latter quantity can be related back to the TOV mass using theoretical models~\citep[e.g.,][]{baiotti17,bauswein20,kashyap22,ecker24,raithel24} to inform the nuclear equation of state. 

We are not the first to suggest combining multiple sub-threshold post-merger gravitational-wave signals, and we won't be the last. \citet{yang18} described two methods; one using the now-classic method of multiplying signal-to-noise Bayes factors, and the other coherently stacking oscillation modes under the relatively strong assumption that the phase of the modes can be accurately determined. Similarly, \cite{Bose2018} describe a method to constrain the neutron star radius by combining total mass measurements from the inspiral phase with the compactness derived from postmerger oscillation frequencies. \citet{criswell23} used empirical relations from numerical-relativity simulations to determine oscillation-mode frequencies from the inspiral parameter estimation, then use BayesWave \citep{BayesWave} to reconstruct the waveforms assuming these mode frequencies are correct. This method was extended to other equation-of-state-proxy parameters in~\citet{Mitra2025}, who studied the systematic biases that can arise from the limitations of empirical relations. Both \citet{criswell23} and \citet{Mitra2025} focussed on constraining the equation of state through indirect measurements of the neutron star radius at some fiducial mass.

In this work, we introduce a different statistical technique that both complements and compliments those of \citet{yang18},~\citet{criswell23} and \citet{Mitra2025}. 
Our statistical framework for combining events is optimal under the assumption that gravitational-wave noise is Gaussian~\citep{smith18}\footnote{In principle, this assumption in the kHz regime of current gravitational-wave observatories is relatively poor, however it can be mitigated effectively using e.g., Allen $\chi^2$ vetos---see~\citet{panther23} for details.}. Using our method, we show that 25-35 events with inspiral signal-to-noise ratio greater than 200 can be combined to give constraints of $\sim 11-20\%$ on the maximum mass of hot, rotating post-merger neutron stars. We discuss the significance of our method as a complement to existing works that constrain the cold nuclear equation of state.

Our Paper is set out as follows. In Sec.~\ref{sec:method}, we outline our methodology, and provide details of the prior choices we make in our analyses. In Sec.~\ref{sec:simulations}, we give details of our synthetic population of gravitational-wave events, including details of the waveforms used for injection and recovery. We present our results in Sec.~\ref{sec:results}, and in Sec.~\ref{sec:discussion} we calculate whether we expect to see a single, loud post-merger remnant before or after we make equation-of-state inferences from our population method (the answer: we're not sure; there's a chance it could be either). We also provide some caveats that could see an acceleration of our method in the coming years, before concluding in the aptly-named "Conclusions" section, Sec.~\ref{sec:conclusions}.

\section{Combining Likelihoods for post-merger signals}\label{sec:method}
We follow the technique outlined in~\citet{smith18} who created \textit{The Bayesian Search} (TBS) to optimally detect a stochastic gravitational-wave background from the inspiral portion of compact binary coalescences. The premise of that work is to perform Bayesian parameter estimation on \textit{every} segment of data, and calculate a fraction $\xi$ of data segments that contain a compact binary coalescense signal versus the total number of segments. In our context, where we search instead for post-merger gravitational waves, we only need to analyse data segments immediately after a binary neutron star inspiral signal is detected. That is, because the post-merger signal is significantly weaker than the inspiral signal, if no inspiral is observed, we can reasonably assume no post-merger signal is present in the data. The denominator of the fraction $\xi$ therefore becomes the total number of binary neutron star mergers for which the inspiral phase is observed in gravitational waves. The numerator of $\xi$ is the number of those that contain a post-merger remnant.

For each data segment $s_i$, a post-merger gravitational-wave strain signal $h_i(\theta_i)$ is present in the data if and only if the system does \textit{not} promptly collapse to form black hole\footnote{A system that promptly collapses will emit gravitational waves through the quasinormal-mode ringing of the black hole, however that will be significantly weaker than neutron-star oscillations, and also at frequencies $\gtrsim6\,{\rm kHz}$~\citep{sarin21}. Given we do not search the data at those frequencies, we can safely say that only data segments for which the systems do not collapse will contain gravitational-wave signals.}. Here, $\theta_i$ are the parameters describing the system, which we discuss below. The likelihood $\likeli{s_i}{\theta_i}$ of the signal given $\theta_i$ is~\citep{veitch15}
\begin{align}
    \ln\likeli{s_i}{\theta_i}\propto\left<s_i-h(\theta_i),\,s_i-h_i(\theta_i)\right>,
    \label{eq:likelihood}
\end{align}
where $h_i(\theta)$ is our post-merger gravitational-wave signal model (discussed below), and the angled brackets denote the usual gravitational-wave inner product assuming Gaussian noise
\begin{align}
    \left<a,b\right>=4{\rm Re}\int_0^\infty df\frac{\tilde{a}(f)\tilde{b}^\star(f)}{S_h(f)}.
\end{align}
Here, an over-tilde represents a Fourier transform, a star the complex conjugate, and $S_h(f)$ is the noise power-spectral density of the gravitational-wave observatory. The likelihood for multiple observatories is simply
\begin{align}
    \likeli{s_i}{\theta_i}\propto\prod_j\likeli{s_i^j}{\theta_i},
\end{align}
where $j$ counts the observatories.\\

Following~\citet{smith18}, we define a mixture model where $\xi$ parameterises whether a post-merger signal exists in the data. That is, for the $i^{\rm th}$ data segment,
\begin{align}
    \likeli{s_i}{\xi,\theta_i}=\xi\likeli{s_i}{\theta_i}+\left(1-\xi\right)\likeli{s_i}{0}.\label{eq:nottheotherlikelihood}
\end{align}
Here, $\likeli{s_i}{0}$ is the likelihood that the data segment does not contain a signal, but only contains Gaussian noise. That is,
\begin{align}
    \likeli{s_i}{0}\propto\left<s_i,\,s_i\right>.
\end{align}

Marginalising Equation~\ref{eq:nottheotherlikelihood} over the parameters $\theta_i$ with a prior $\pi(\theta_i)$ gives the marginal likelihood 
\begin{align}
    \likeli{s_i}{\xi}=\xi\mathcal{Z}_s^i+\left(1-\xi\right)\mathcal{Z}_n^{i}.
\end{align}
Here, $\mathcal{Z}_s^i$ and $\mathcal{Z}_n^i$ are respectively the signal and noise evidences given by
\begin{align}\label{eq:evidence}
    \mathcal{Z}_s^i=&\int d\theta\likeli{s_i}{\theta_i}\pi(\theta_i),\\
    \mathcal{Z}_n^i=&\likeli{s_i}{0}.
\end{align}
For a set of $N$ mergers, the total likelihood is
\begin{align}
    \likeli{\{s_i\}}{\xi}=\prod_{i=0}^N\likeli{s_i}{\xi}.
\end{align}

Bayes theorem implies the posterior probability of the fraction of data segments that include a post-merger gravitational-wave signal is given by
\begin{align}
    p\left(\xi|\{s_i\}\right)\propto\likeli{\{s_i\}}{\xi}\pi(\xi),
    \label{eq:XiPosterior}
\end{align}
where $\pi(\xi)$ is our prior belief on the fraction of mergers $\xi$ that leave behind a gravitational-wave-emitting post-merger remnant. Astrophysically, $\pi(\xi)$ is therefore our prior belief for the fraction of events that we expect to not collapse promptly divided by the total number of events.

\begin{figure}
    \centering
    \includegraphics[width=1.0\linewidth]{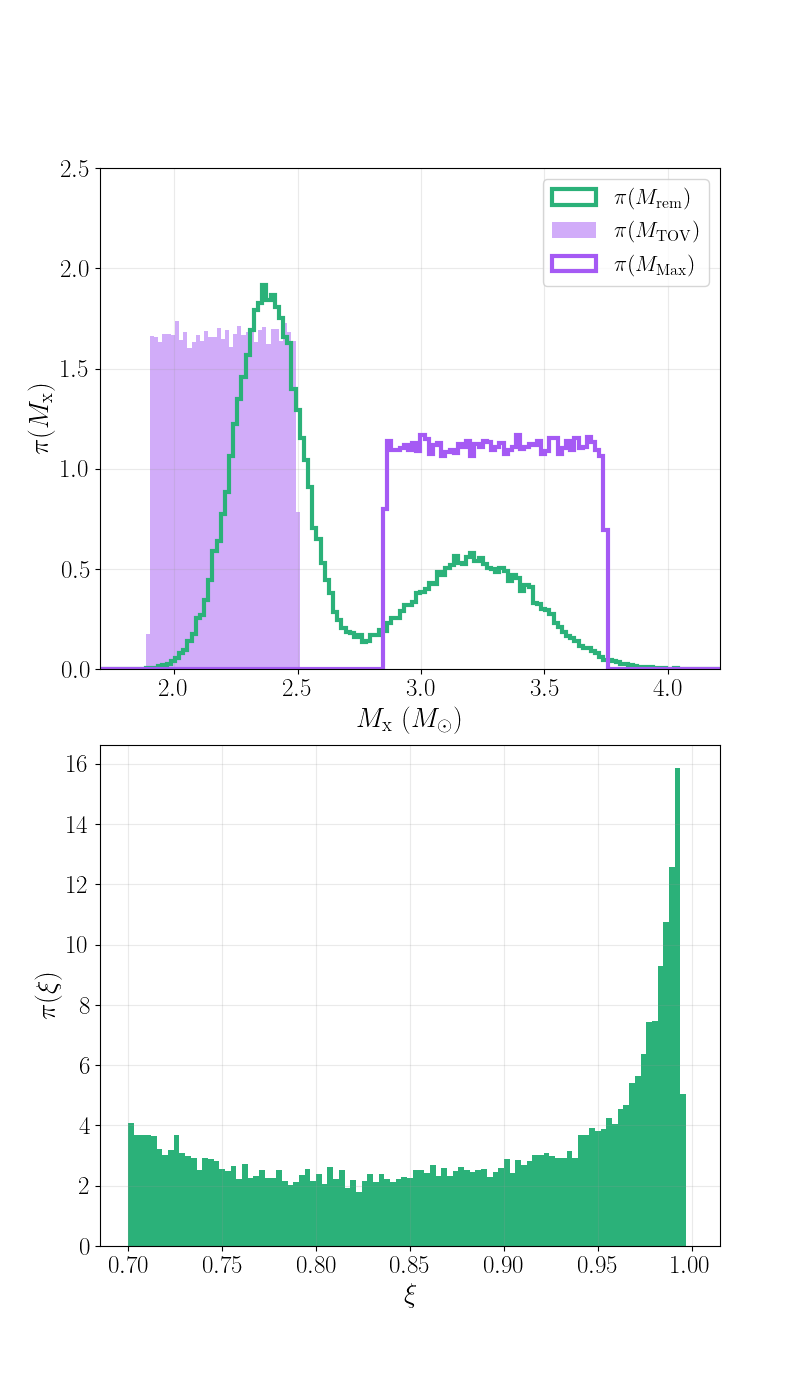}
    \caption{Prior probability distributions used in this work. We assume a bimodal remnant mass distribution calculated from the expected distribution of merging binary neutron star pairs with masses drawn independently from the observed neutron star population (top panel; green histogram). We choose uniform priors on $M_\mathrm{TOV}$ (top panel; purple filled histogram) and $M_\mathrm{Max} = 1.5M_\mathrm{TOV}$ (top panel; purple hollow histogram). The bottom panel shows the resultant $\xi$ prior for a given fixed value of $M_\mathrm{Max}$,calculated using Eq. \ref{eq:XiPrior}. There is a lower and upper bound on the prior distribution of $\xi$ as some of the neutron star remnant masses always lie above or below the limits our prior on $M_\mathrm{Max}$.}
    \label{fig:priors}
\end{figure}

Suppose we have an expected probability distribution for the \textit{gravitational} remnant masses of our binary neutron star mergers, $\pi(M_{\rm rem})$. The term $\xi$ represents the fraction of this distribution that does not exceed the maximum neutron star mass $M_{\rm Max}$ . This can be calculated as a normalised integral of $\pi(M_\mathrm{rem})$ up to the maximum mass $M_{\rm max}$, i.e., 
\begin{align}
    \xi=\frac{\int_0^{M_{\rm max}}dM_{\rm rem}\pi(M_{\rm rem})}{\int_0^\infty dM_{\rm rem}\pi(M_{\rm rem})}.\label{eq:IndividualXi}
\end{align}
In general, $M_{\rm max}$ can be related to the TOV mass $M_{\rm TOV}$ as $M_{\rm max}=\chi M_{\rm TOV}$, where $\chi$ is an equation-of-state-dependent parameter~\citep[see][and references therein]{sarin21}. Given we do not know $M_{\rm TOV}$, we choose to express our prior belief in the true value of $M_\mathrm{TOV}$ as $\pi(M_{\rm TOV}) = \mathcal{U}(1.9\,M_\odot, 2.5\,M_\odot)$, where $\mathcal{U}(a, b)$ is a uniform distribution bounded by $a$ and $b$. Therefore, our prior on $M_{\rm max}$ becomes
\begin{align}
    \pi(M_{\rm max})=\chi\pi(M_{\rm TOV})=\chi\mathcal{U}(1.9\,M_\odot, 2.5\,M_\odot).
\end{align}
These prior distributions for $\pi(M_{\rm max})$ and $\pi(M_{\rm TOV})$ are shown in the top panel of Figure~\ref{fig:priors} for $\chi=1.5$ as the empty-purple and shaded-purple histograms, respectively. In principle one could derive a multi-messenger prior on $M_\mathrm{TOV}$ from the wealth of electromagnetic and inspiral GW constraints presently available, however we choose a more agnostic prior in this work. In this work, we fix the value of $\chi$ as the goal of this paper is as a proof-of-concept. In practice, one should use equation-of-state dependent values of $\chi$, either inferred empirically through observations or calculated based on numerical-relativity simulations, where the equation-of-state dependent relationship between $M_\mathrm{TOV}$ and $M_\mathrm{Max}$ was first demonstrated \citep{Bauswein2013}. In general, $M_\mathrm{Max}$ and hence $\chi$ depend on the rotational profile of the remnant as well as the equation-of-state \citep{Iosif2022, Cassing2024, Musolino2024}. Simulations imply $\chi$ is equation-of-state dependent, but between 1.3 and 1.6~\citep{shibata00,shibata06,baiotti17,agathos20}, however there are not yet enough numerical-relativity simulations that incorporate rotational effects to cover the full range of such a prior. This is a potential source of uncertainty that we are not taking into account in our subsequent demonstration: when applied to real data, we expect our final posterior on $M_\mathrm{Max}$ to be broader if one incorporates a broader prior on $\chi$. We leave empirical inference of $\chi$ as a free parameter of our mixture model to a future work.

It is important to carefully define $M_\mathrm{rem}$. Values of $\chi$ above are calculated by evaluating equilibrium sequences of isolated neutron stars with various rotation profiles. There, $M_\mathrm{Max}$ is the \textit{gravitational} mass at which the neutron star becomes unstable. However, the total mass $M_\mathrm{Tot}$ is the sum of the gravitational masses of the binary neutron star components at finite orbital separation. This gravitational mass is not conserved in the merger. We calculate the final remnant mass by using conservation of \textit{rest} mass, other than mass lost to dynamic and thermal effects, which numerical simulations show to be relatively small~\citep[e.g.,][]{, Radice2016}. We use the approximate conversion between gravitational and rest masses of $M^{\rm rest}=M+0.075M^2$ \citep{timmes96}. The resultant remnant mass distribution for our prior $\pi(M_{\rm rem})$ is shown in green in the top panel of Figure~\ref{fig:priors}.

We can now re-express Equation~\ref{eq:IndividualXi} as a prior probability distribution on $\xi$ as
\begin{align}
    \pi(\xi)=\frac{\int_0^{\pi(M_{\rm max})}dM_{\rm rem}\pi(M_{\rm rem})}{\int_0^\infty dM_{\rm rem}\pi(M_{\rm rem})}.
    \label{eq:XiPrior}
\end{align}
This can be calculated numerically by drawing samples from $\pi(M_{\rm max})$ and, for each sample, integrating $\pi(M_{\rm rem})$ to evaluate a single sample in $\xi$. 

Our final ingredient to evaluating Equation~\ref{eq:XiPrior}, and hence the posterior on $\xi$ through Equation~\ref{eq:XiPosterior}, is to determine a prior for $M_{\rm rem}$, the binary neutron star remnant mass. Suppose we knew the distribution of progenitor masses $\pi(m_1,\,m_2)$, where $m_1>m_2$ are the neutron star component progenitor masses. For the purposes of this work we assume the masses of each progenitor are independent from one another, and that the individual progenitor mass follows a Gaussian mixture model
\begin{align}
    \pi(M_a) = (1-\epsilon) \mathcal{N}(\mu_1,\,\sigma_1)+\epsilon\mathcal{N}(\mu_2,\,\sigma_2).
\end{align}
where $a$ labels the progenitor star component and $\mathcal{N}(\mu,\,\sigma)$ is a normal distribution of mean $\mu$ and standard deviation $\sigma$. We let $(\mu_1,\,\sigma_1)=(1.31,0.11)\,M_\odot$ from the Galactic neutron star mass distribution~\citep{Kiziltan2013, Antonadis2016}, $(\mu_2,\,\sigma_2)=(1.80,\,0.21)\,M_\odot$ consistent with the mass of GW190425 \citep{GW190425}, and $\epsilon=0.35$ following~\citet{sarin20}. In principle, one could also marginalise over these parameters to simultaneously determine the mass distribution of the progenitors. We justify holding these fixed because we anticipate the binary neutron star mass distribution to be (somewhat) well-measured from neutron star inspirals by the time we can statistically measure the post-merger signal using our method. For example, below we show simulations of 70 synthetic post-merger events, each of which will have their progenitor masses measured from the inspiral gravitational-wave signals, and a subsequesnt population distribution will be inferrable~\citep{Read2021}. 

We do not enforce any prior on the mass ratio $q = m_2/m_1$. It has been shown that $M_\mathrm{Max}$ is rather sensitive to $q$: For values of $q\sim0.7$, $M_\mathrm{Max}$ can be reduced by around $0.1\,\mathrm{M_\odot}$ \citep{bauswein20, Bauswein2021}. This effect can be incorporated into the analysis method by taking the mass ratio measured from the inspiral and converting this into a prior on the mass loss due to the effect of the mass ratio. We currently fix the value for the mass loss in our calculation of $M_\mathrm{rem}$, but a mass-ratio-dependent prior can be used instead. We note that this effect may be important as it could induce a systematic bias in the measurement of $M_\mathrm{Max}$ in a population of events with $q\neq1$, however as this work is a first proof-of-concept, we defer an exploration of this effect to future study.

The bottom panel of Figure~\ref{fig:priors} shows our resultant prior for $\xi$ using Equation~\ref{eq:XiPrior} and the priors shown in the top panel. We note the lower and upper bounds on $\pi(\xi)$ arise from the remnant masses always being below or above the limits of our prior on $M_{\rm max}$. $\pi(\xi)$ is strongly peaked at $\xi\sim1$. By considering the top panel of \ref{fig:priors} we can understand the qualitative shape of $\pi(\xi)$: only a small fraction of the remnant mass distribution lies above the upper bound on $M_\mathrm{Max}$, therefore our prior belief is that the majority of merging systems we simulate will result in a remnant that is (quasi-)stable. We find that although our prior on the remnant mass distribution is highly informed by the masses of GW190425, changing these priors does not have a substantial impact on the overall recovery of the posterior on $M_\mathrm{Max}$.

\section{Simulations}\label{sec:simulations}
To date, only two confident gravitational-wave detections of binary neutron stars have been made~\citep[although a sub-threshold candidate has been identified by][]{Niu2025}. Attempts to infer the presence of any post-merger gravitational-wave signal in the moments following these events overwhelmingly favour the noise hypothesis owing to the relatively poor kHz sensitivity of the current generation of observatories~\citep{GW170817postmerger, GW170817postlong, GW190425, Abchouyeh2023, Grace2024}. Therefore, we validate our method with a series of simulations: we inject post-merger waveforms into segments of Gaussian noise coloured by the detector sensitivities we expect for future detector networks. Namely, we use a `third-generation' (3G) detector network consisting of two Cosmic Explorer~\citep[CE;][]{CosmicExplorer, CEReitze} interferometers, noting that detection efficiencies with networks including the Einstein Telescope~\citep{EinsteinTelescope} or NEMO~\citep{ackley20} would result in quantitatively similar outcomes to those presented below. Although confident detections have been predicted with 3G networks, the vast majority of post-merger signals will remain below the established threshold for a confident detection. 

\begin{figure}
    \centering
    \includegraphics[width=1.0\linewidth]{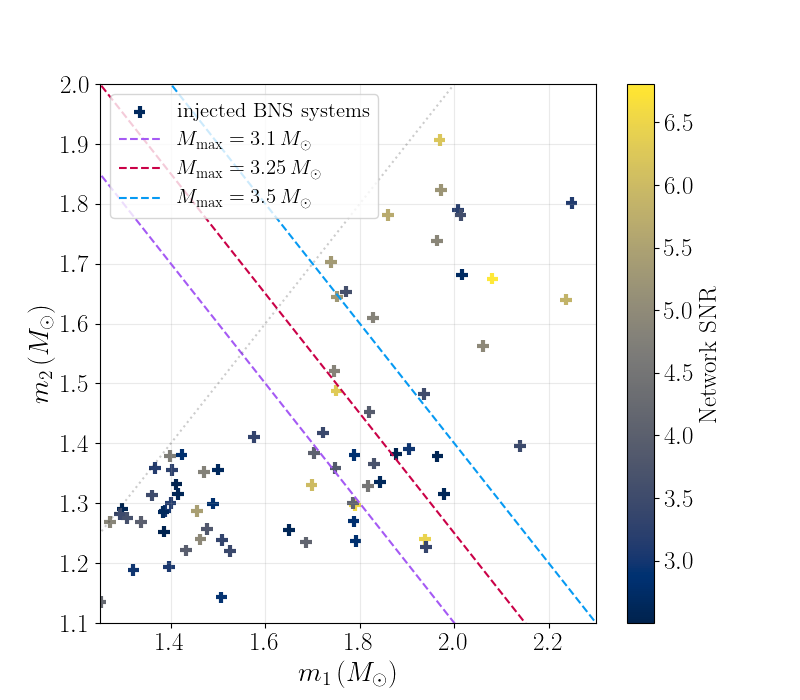}
    \caption{Masses of simulated binary neutron star signals used in this work. Dashed lines show different values of $M_\mathrm{Max}$ used to evaluate the robustness of the technique. The dotted line shows $m_1 = m_2$. The colourbar indicates the network SNR of the injected signals. For each $M_\mathrm{Max} = [3.1, 3.25, 3.5]\,\mathrm{M_\odot}$, the fraction of data segments that contain injected signals is $\xi = [0.56, 0.70, 0.84]$.}
    \label{fig:masses}
\end{figure}

\begin{figure}
    \centering
    \includegraphics[width=1.0\linewidth]{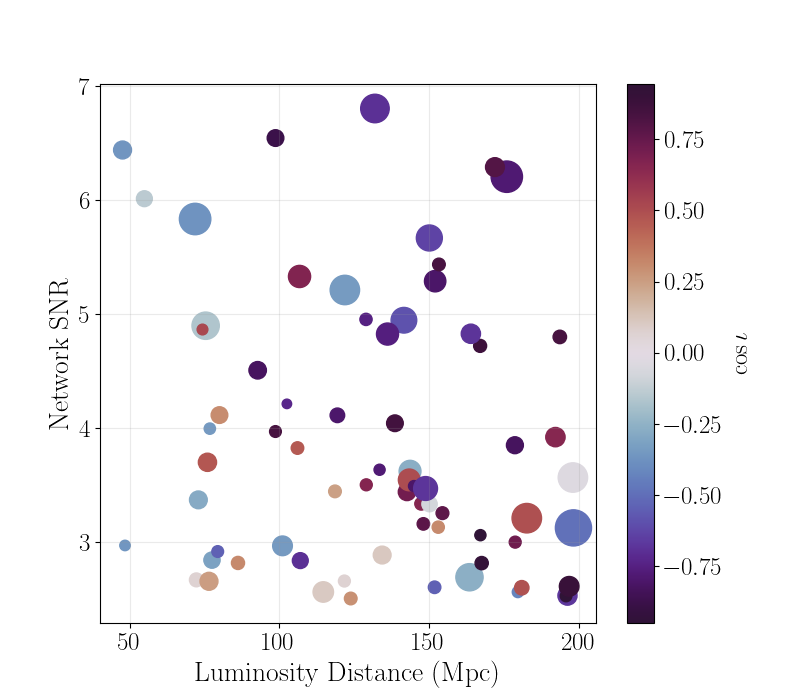}
    \caption{Distribution of luminosity distances and network postmerger SNRs for our injected signal population. Luminosity distances are drawn such that they are uniform in co-moving volume. The size of each point indicates the relative total mass of each binary neutron star system. More massive systems, and face-on systems ($\cos\iota\simeq\pm1$, represented by higher color saturations) produce signals with larger amplitudes, allowing them to attain considerable SNR out to larger luminosity distances.}
    \label{fig:snrdl}
\end{figure}

We generate an injection set comprising 70 binary neutron star post-merger remnant signals with network $\mathrm{SNR}\in[2.5,7]$, representing a population of signals that lie below the expected confident detection threshold in Gaussian noise colored by the noise power spectral density of Cosmic Explorer. Post-merger waveforms are computed using the \texttt{NRPM\_only} model of \cite{Breschi2024}, which is a frequency-domain analytic model calibrated to numerical-relativity simulations. Component masses $m_1$ and $m_2$ are drawn independently from the bimodal mass distribution described above. For each pair of values we define $m_1\geq m_2$; the component masses of our simulated binaries are shown in Fig. \ref{fig:masses}, where the colour of the marker reflects the network SNR. All other waveform parameters are drawn from a uniform distribution with the exception of the luminosity distances, which are sampled to be uniform in co-moving volume up to $200\,\mathrm{Mpc}$ using a Planck 2018 cosmology~\citep{Planck}. Figure~\ref{fig:snrdl} shows the SNR distribution of our injected signals as a function of their luminosity distance, coloured by the inclination angle $\iota$, and where the size of the point represents the relative total mass of the binary. Each of these parameters affects the amplitude of the emitted post-merger gravitational-wave signal, and hence the SNR: more massive systems, those that are close to face-on, and those that are nearby, will have larger amplitudes. We briefly discuss the potential consequences of these effects in Section \ref{sec:discussion}.

We simulate three different maximum neutron star mass scenarios given by $M_\mathrm{Max} = [3.1, 3.25, 3.4]\,\mathrm{M_\odot}$. For each threshold $M_\mathrm{Max}$, if $M_\mathrm{Tot}\leq M_\mathrm{Max}$, our simulated waveforms are injected into one-second segments of coloured Gaussian noise representing observations made by our two-detector network. For events where $M_\mathrm{Tot}>M_\mathrm{Max}$, no signal is injected into our coloured Gaussian noise as we assume the system collapses promptly to form a black hole, which emits relatively weak gravitational waves outside the band of our observatories. Bayesian inference can be used to calculate signal and noise evidences (Eq. \ref{eq:evidence}) for each segment, including those for which no signal is injected, by marginalizing over all the parameters of the waveform model. 

The fraction of data segments that do contain signals for each of our $M_\mathrm{Max} = [3.1, 3.3, 3.5]\,\mathrm{M_\odot}$ is $\xi = [0.56, 0.78, 0.84]$, respectively. In the case of $M_\mathrm{Max}=3.1\,\mathrm{M_\odot}$, it should be noted that the fraction of data segments expected to contain signals lies outside the domain of compact support of $\xi$ due to the finite sample size. This is likely to lead to railing of the posterior. As can be seen in Fig. \ref{fig:masses}, there are subpopulations that always produce a merger remnant signal (below $3.1\,\mathrm{M_\odot}$) or never produce a post-merger remnant signal (above $3.5\,\mathrm{M_\odot}$) that are common to each injection set.

\subsection{Signal injection and recovery}
We use the \texttt{Bilby} inference package~\citep{ashton19,romeroshaw20} with the \texttt{dynesty} sampler~\citep{speagle20} to calculate posterior distributions on each of the free parameters of the \texttt{NRPMw\_pmonly} post-merger waveform model~\citep{Breschi2024}. As \texttt{dynesty} is a nested sampler~\citep{feroz08}, this also calculates our signal and noise evidences.\\

We require a prior probability distribution for each parameter in our model. At CE sensitivity, it will be possible to reliably detect and determine the time of coalescence of the inspiralling binary neutron stars with high precision, as the typical network SNR for the inspiral signal associated with our post-merger remanants is $\mathrm{SNR}\sim 300$ (see below). Methods to estimate the physical parameters of these systems are well-established, including component masses, luminosity distance and sky localization~\cite[e.g., see][]{smith21}. As our chosen post-merger-waveform model is parameterised by the properties of the inspiralling neutron stars, we could opt to impose prior constraints on all our post-merger waveform parameters informed by inference on the inspiral signal. However, we choose only to impose such `inspiral-informed priors' on the extrinsic parameters of the post-merger signal. 

\begin{figure}
    \centering
    \includegraphics[width=1.0\linewidth]{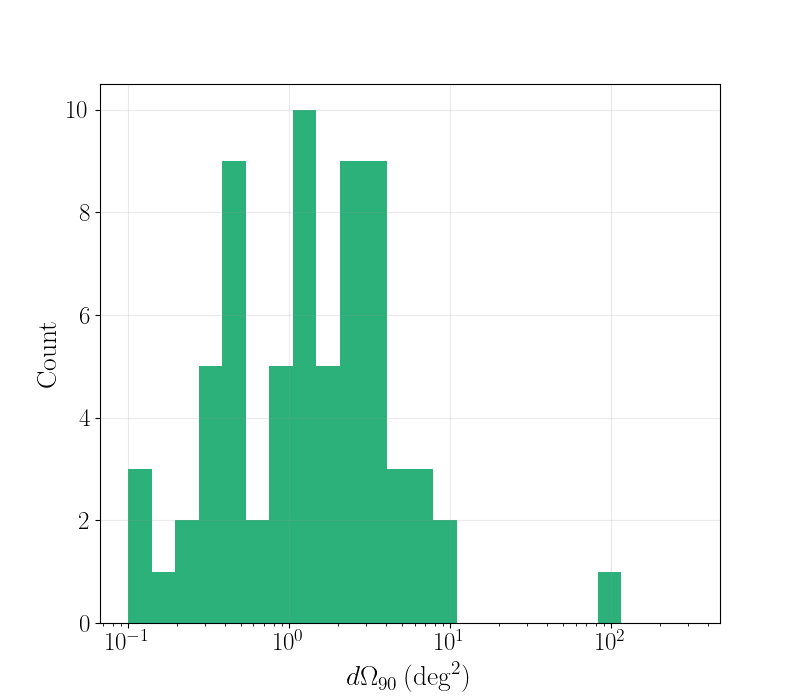}
    \caption{Distribution of $90\%$ credible-interval sky-localization uncertainties $d\Omega_{90}$ computed for the inspiral portion of our signals. The majority of signals are localized to under $5\deg^2$, justifying our use of delta-function priors for our post-merger analyses.}
    \label{fig:fisher}
\end{figure}

In accordance with the above argument, we impose delta-function-prior constraints on the three-dimensional localization of our injected binary neutron star signals. The sensitivity of future detectors means the anticipated sky localization volumes for our events are small. To test the validity of this assumption, we use \texttt{gwbench}~\citep{gwbench} to calculate the median 90 per cent credible interval sky localization uncertainty of the inspiral binary neutron star signals via the Fisher Information Formalism. We use the \texttt{tf2\_tidal} waveform model to generate representative inspiral waveforms for the injection parameters drawn above, and inject them into colored Gaussian noise drawn from the same power spectral density as used in the post-merger analysis. The SNR of these inspiral signals ranges from $\rho_\mathrm{net} = 200 - 500$. 
We then use \texttt{gwbench} to estimate the lower bound on the accuracy with which the different parameters of the model---including three-dimensional sky localization---can be recovered. We find a median localization uncertainty of $d\Omega_\mathrm{90}= 1.69\deg^2$ and the typical $1\sigma$ uncertainty in the luminosity distance to be $\sigma_{d_L} =1.25\,\mathrm{Mpc}$, assuming a single-mode Gaussian posterior on $d_L$; see Fig.~\ref{fig:fisher}. This validates our choice of delta-function priors for these parameters\footnote{We note a single outlier, where the sky localization of $\sim 100\deg^2$ arises due to the sub-optimal position of the signal relative to the sensitivity of the two-detector network across the sky.}, and has the added benefit of providing substantive speed-up of our post-merger inference calculations.

We could also choose to use inspiral-informed priors for the masses in our post-merger waveform model. While this would undoubtedly improve the constraints we would achieve, we believe it to be an overly optimistic assumption, because it implicitly assumes that the post-merger waveforms are a faithful representation of our real-world expectations. However, using the inspiral-informed priors would implicitly assume that e.g., there are no temperature-induced phase transitions that could induce strange features in the post-merger signal morphology that may manifest as different mass inferences. Such phase transitions may result in the remnant having a different equation-of-state to that describing the inspiralling neutron stars, and could have a substantial impact on the post-merger GW emission \citep{Oechslin2007, Blacker2023}. To that end, we consider this a conservative assumption.

We sample the binary neutron star masses in the native parameter space of the \texttt{NRPMw\_pmonly} model $\{M_\mathrm{tot}, q = m_1/m_2\}$ where 
\begin{align}
    \pi(M_\mathrm{tot}) &= \mathcal{U}(2\,M_{\sun} , 5\,M_{\sun});\,\pi(q) = \mathcal{U}(1, 1.5).
\end{align}
Note the different definition of $q$ compared to the standard definition used in \texttt{Bilby}; i.e., $q_\mathrm{bilby} = 1/q$. We set a generous upper limit on $M_\mathrm{Tot}$ to prevent parameter railing. We also impose prior constraints such that
\begin{align}
    m_1, m_2 &\in [1\,M_{\sun}, 2.5\,M_{\sun}].
\end{align}
These mass priors are motivated by our understanding of the neutron star mass distribution as described above. We only consider non-precessing inspirals, with low spins, and hence choose our spin priors as 
\begin{equation}
    \pi(s_\mathrm{iz}) = \mathcal{U}(0, 0.2),\,i\in\{1, 2\}.
\end{equation}
Progenitor neutron star spins have minimal impact on post-merger gravitational-wave emission unless the pre-merger neutron stars have spin periods of $\sim\mathrm{ms}$, which is expected in only $\sim 4\%$ of systems~\citep{Rosswog2023}.

We assume tidal deformabilites are uniformly distributed and sampled in $\Lambda_i$,
\begin{equation}
    \pi(\Lambda_i) = \mathcal{U}(10, 3000),\,i\in\{1, 2\}.
\end{equation}
We impose a lower bound of $\Lambda_i = 100$ to ensure that waveforms can be generated across the whole prior volume as at low masses the model fails for $\Lambda_i < 100$.

Our prior on the binary inclination angle $\iota$ is uniform in $\cos(\iota) \in [-1, 1]$. We use delta-function priors on the inspiral coalescence phase $\phi_c$ as it has no impact on the \texttt{NRPMw\_pmonly} model, which comprises only the post-merger portion of the signal. We also fix the merger time, which is known \textit{a priori} from the inspiral, and the polarization angle to improve the efficiency of the computation.

The waveform model \texttt{NRPMw\_pmonly} requires three additional post-merger parameters: the mass-weighted post-merger time of collapse $t_\mathrm{coll}/M$, the mass-weighted post-merger frequency drift $\alpha/M^2$ and the post-merger phase offset from the coalescence phase $\phi_\mathrm{pm}$. We fix $t_\mathrm{coll}/M=700$ to represent that we expect to observe signals that have durations of $\gtrsim 25\,\mathrm{ms}$. Moreover, several works have found little information to discriminate the true value of this parameter will be recoverable~\citep{easter21,dhani24}. Our prior on the frequency drift is uniform in $\alpha/M^2\in[-10^{-4}, 10^{-4}]$ and post-merger phase uniform and periodic in $\phi_\mathrm{pm}\in[0, 2\pi]$.

\section{Results}\label{sec:results}
\begin{figure}
    \centering
    \includegraphics[width=1.0\linewidth]{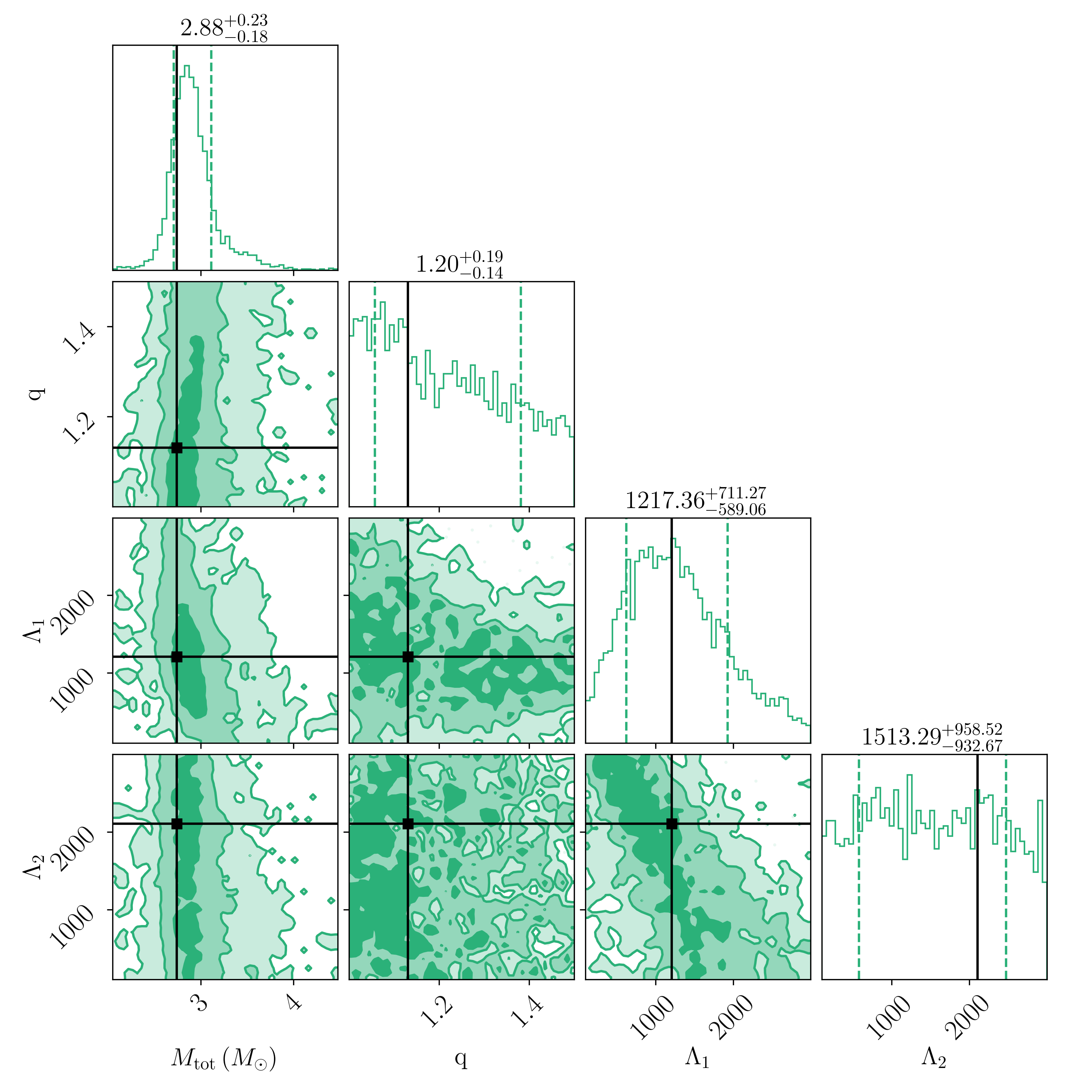}
    \caption{Posterior probability distributions on $M_\mathrm{tot}$, mass ratio $q = m_1/m_2$ and tidal deformabilities $\Lambda_1, \Lambda_2$ for a post-merger signal with $\mathrm{SNR=3.7}$. In this case, $M_\mathrm{tot}$ can be recovered well, however the posterior distributions of other parameters do not have strong predictive power.}
    \label{fig:corner}
\end{figure}

Because our individual injections have small post-merger SNRs, the posterior probability distribution on the masses of each individual post-merger system are not generally informative. As an illustrative example, we show the posterior probability distributions (Fig. \ref{fig:corner}) recovered for one of our injections where we find there is moderate-to-strong evidence for the signal hypothesis, with $\ln\mathrm{BF} = 3.4$. This represents one of the most confident detections in our sample. Although the mass posterior is tightly peaked close to the injected $M_\mathrm{Tot}$ value, the distributions on the mass ratio $q$ and the tidal deformabilities $\Lambda_{1,2}$ are uninformative. Based on the network SNR of the signal, $\rho_\mathrm{net}= 3.7$, such an event would be difficult to distinguish from noise processes in either Gaussian or real detector noise.

To determine how a population of binary neutron star events can allow us to constrain $M_\mathrm{Max}$, we combine the signal and noise evidences from our population of $N = [5, 25, 45, 65, 70]$ binary neutron star mergers described in Section\ref{sec:method} to obtain a posterior probability distribution for $M_\mathrm{Max}$. For each value of $M_\mathrm{Max, True} \in \{3.1, 3.25, 3.4\}\,M_\odot$, signals are injected into Gaussian noise if $M_\mathrm{Tot, i}<M_\mathrm{Max, True}$.

\begin{figure*}
    \centering
    \includegraphics[width=\textwidth]{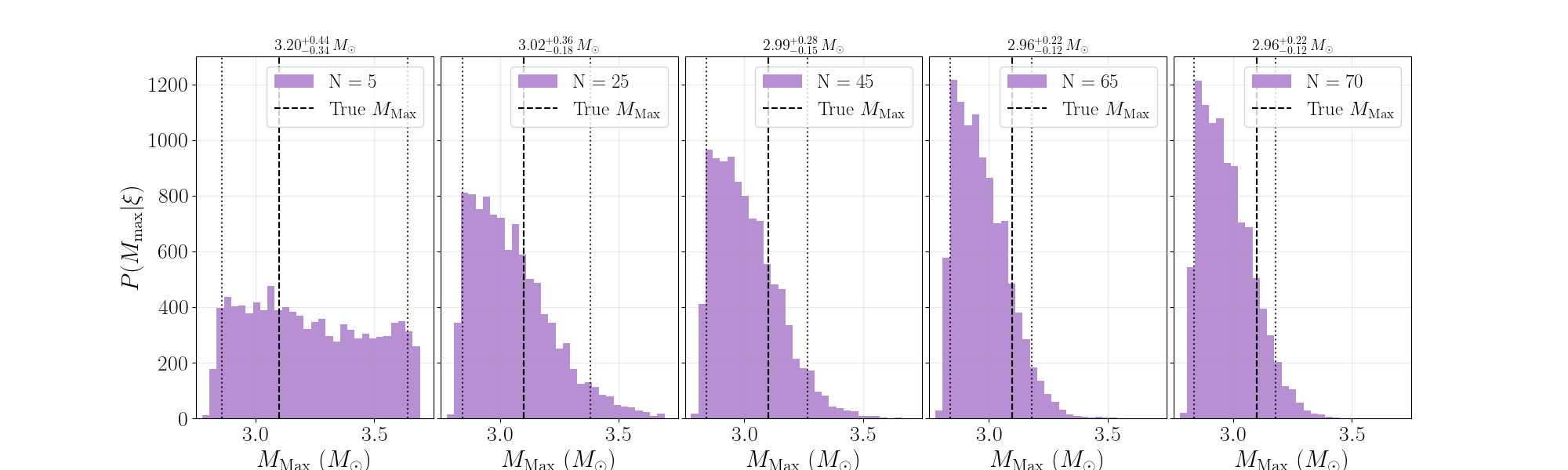}
    \caption{Recovered normalized posterior distributions for $M_\mathrm{Max}$ with an increasing number of injected signals. Here,  $M_\mathrm{Max, True}=3.1\,\mathrm{M_\odot}$ (vertical dashed line). Vertical dotted lines indicate the upper and lower bounds of the 90\% confidence interval on $M_\mathrm{max}$.}
    \label{fig:mmax31}
\end{figure*}

In Fig. \ref{fig:mmax31}, we show the posterior distributions for $M_{\rm Max}$ for the case of $M_{\rm tot}=3.1\,M_{\odot}$, with the panels going from left to right showing $N=5$, 25, 45, 65, and 70 events, respectively. For example, the second-left-most panel shows that for 25 events we recover the injected value of $M_\mathrm{Max}$ (vertical dashed line) within the 90\% credible interval (vertical dotted lines). There is some railing of the posterior at low $M_{\rm Max}$: this is physical and expected when we consider our injected population. The fraction of events in our sample with $M_\mathrm{Tot}<3.1\,\mathrm{M_\odot}$ into which a signal is injected is $\xi_\mathrm{injected} = 0.56$. This is due to our finite sample size---with a larger sample size, this value should converge toward the theoretical expectation of $\xi_\mathrm{True}$. Nevertheless, the true value of $M_\mathrm{Max}$ lies well within the 90\% confidence interval of the posterior distribution. With 70  events (right-most panel), we find $M_\mathrm{Max} = 2.96^{+0.23}_{-0.12}\,\mathrm{M_\odot}$, where the 90\% confidence interval corresponds to a $\sim 12\%$ uncertainty in the value of $M_\mathrm{Max}$.

\begin{figure*}
    \centering
    \includegraphics[width=\textwidth]{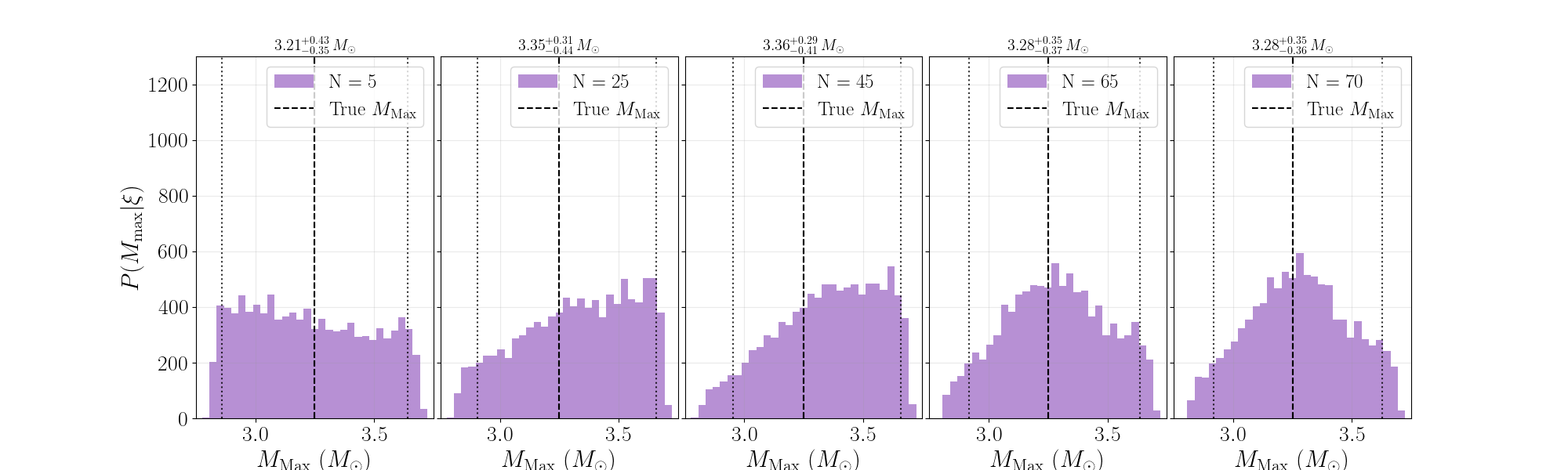}
    \caption{Same as Fig. \ref{fig:mmax31} for $M_\mathrm{tot}<3.25\,\mathrm{M_\odot}$}
    \label{fig:mmax33}
\end{figure*}

\begin{figure*}
    \centering
    \includegraphics[width=\textwidth]{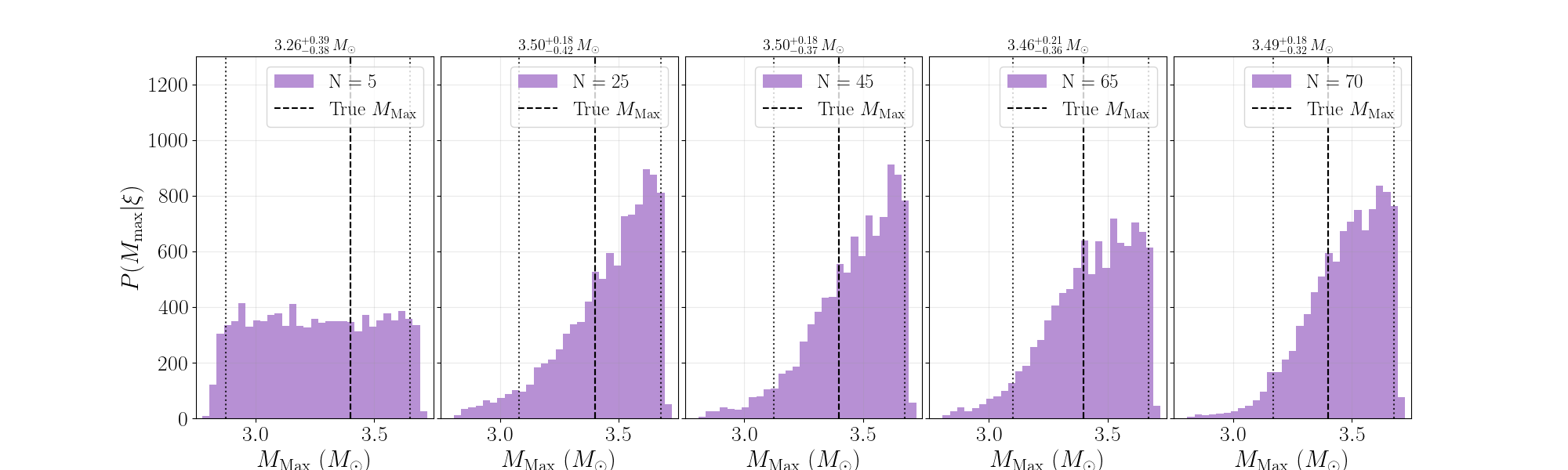}
    \caption{Same as Fig. \ref{fig:mmax31} for $M_\mathrm{tot}<3.4\,\mathrm{M_\odot}$}
    \label{fig:mmax35}
\end{figure*}

The same posterior probabilities for $M_\mathrm{Max} = 3.25\,\mathrm{M_\odot}$ and $M_\mathrm{Max} = 3.4\,\mathrm{M_\odot}$ are shown in Figs. \ref{fig:mmax33} and \ref{fig:mmax35}, respectively. With $N\sim35$ events, our method can accurately recover $M_\mathrm{Max}$ to within $10-20\%$. 

\begin{figure*}
    \centering
    \includegraphics[width=1.0\linewidth]{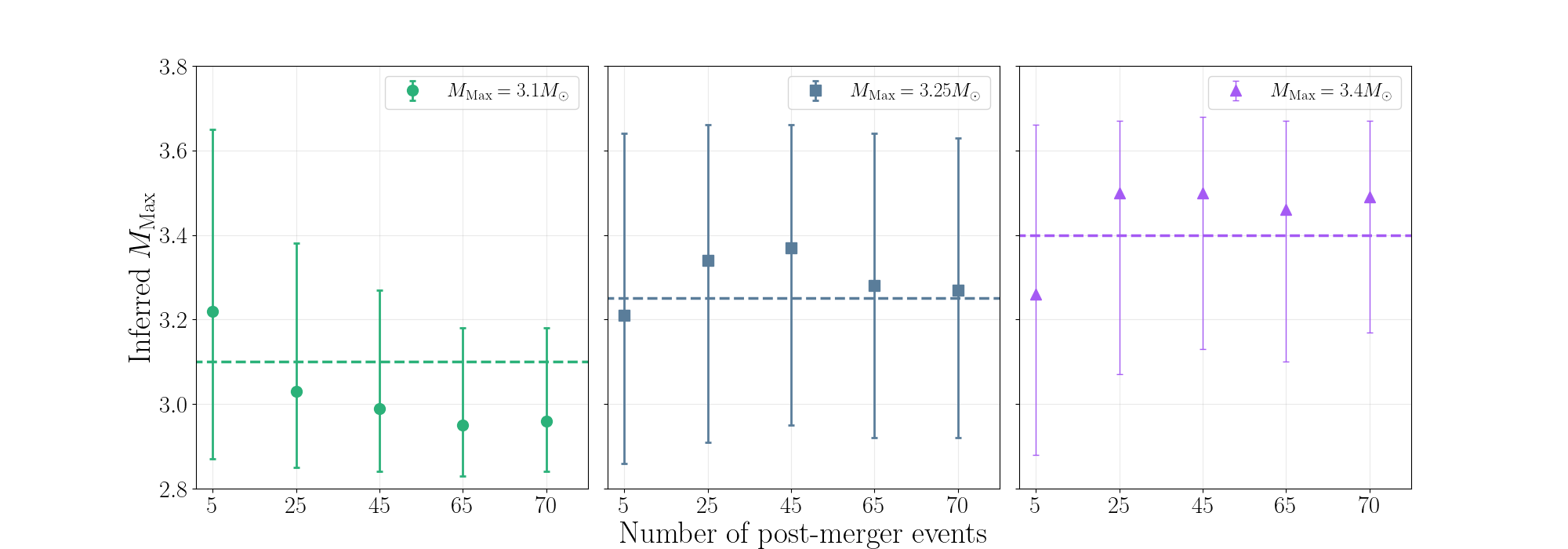}
    \caption{Recovered median $M_\mathrm{Max}$ for each experiment. Error bars indicate 90\% CI. With 70 events, the percentage error in estimation of $M_\mathrm{Max}$ is 11, 21, and 15\% respectively for $M_\mathrm{Max, true} = 3.1, 3.25, 3.4\,\mathrm{M_\odot}$}
    \label{fig:medians}
\end{figure*}

A summary of these posterior distributions is shown in Figure \ref{fig:medians}, where we see that we tend to overestimate (underestimate) the median $M_\mathrm{Max}$ for true $M_\mathrm{Max}$ values that lie at the high (low) mass end of the domain of compact support for $M_\mathrm{Max}$. The underestimation of $M_\mathrm{Max}$ in the $M_\mathrm{Max, True} = 3.1\,M_\odot$ case is driven by the finite sample size having $\xi_\mathrm{inject}<0.7$, however the overestimation of $M_\mathrm{Max}$ in the $M_\mathrm{Max, True} = 3.4\,M_\odot$ case is driven by several of our data segments where no signal is injected returning $\ln\mathrm{BF}\sim -0.5-1$. This results in the likelihood being skewed toward larger values of $\xi$ as there is insufficient events that are conclusively identified as noise with $\ln\mathrm{BF}<-0.5$, and because our prior is heavily weighted toward large values of $\xi$ due to the majority of the $M_\mathrm{rem}$ distribution lying below $3.4\,M_\odot$. We discuss the impact of this bias and potential mitigation in the next section. Nevertheless, we find that in all cases, we accurately estimate $M_\mathrm{Max}$ with a precision of $10-20\%$. This corresponds to a measurement of $M_\mathrm{TOV}$ with a precision of $12-21\%$ for a fixed value of $\chi = 1.5$. This represents an optimistic estimate, as incorporating full physical priors for the mass-ratio-dependent mass loss during merger and the possible range of values for $\chi$ will lead to information loss. However, in the event these values can be well constrained either empirically or from numerical-relativity simulations, such a measurement can effectively be used as an indirect probe of phase transitions \citep[similar to that described in][]{Bauswein2019} by comparing $M_\mathrm{Max}$ and $M_\mathrm{TOV}$ measured from the inspiralling neutron star population and the values obtained from our method that is sensitive to the configuration of the newly formed hypermassive merger remnant.

\section{Single event or population?}\label{sec:discussion}

The sensitivity of the posterior on $\xi$, and hence $M_\mathrm{Max}$, to over-fitting of the noise represents an unfortunate limitation of our method. This sensitivity to the noise realisations reflects the nature of the post-merger waveforms, which have broad frequency domain features. 
This is particularly challenging for the low-SNR signals we are deliberately targeting here; post-merger waveforms typically have one dominant spectral maximum, implying they resemble simple but non-stationary noise artifacts. Without the secondary or tertiary maxima in their frequency spectrum that are usually well below the noise floor, the waveforms can inadvertently fit generic noise features.

\begin{figure}
    \centering
    \includegraphics[width=1.0\linewidth]{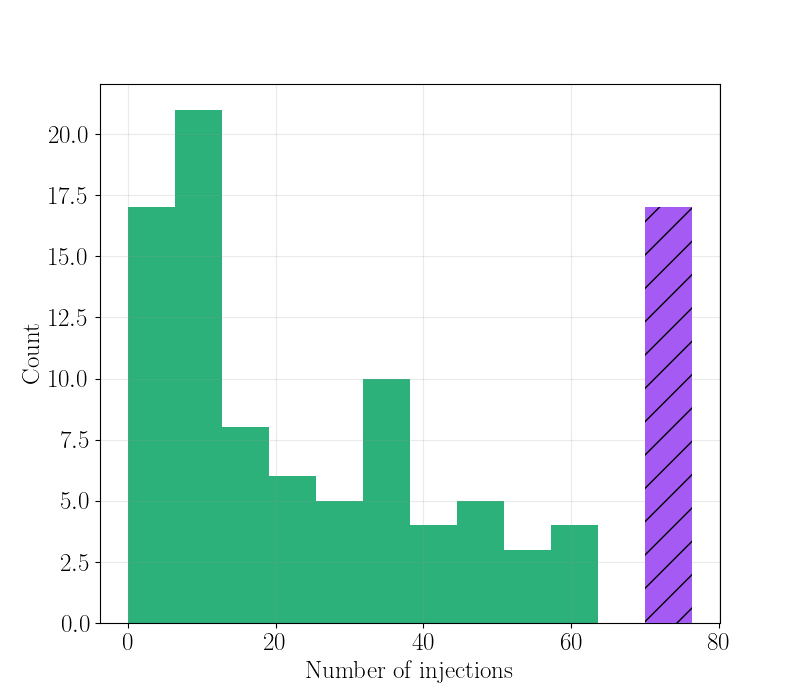}
    \caption{Number of simulated signals that are added to the injection set before a signal with SNR$>12$ is detected for binary neutron star systems that are distributed uniform in co-moving volume out to $d_L=200\,\mathrm{Mpc}$. In 17\% of our realizations, no signal with SNR$>12$ is generated before the injection set is complete (violet hatched bin).}
    \label{fig:timetoloud}
\end{figure}

Due to the limitations of this method, it is prudent to consider whether it is likely we will detect a single binary neutron star post-merger remnant before the $25-35$ events needed to constrain $M_\mathrm{Max}$ via the population approach. To investigate this question, we generate 100 realizations of our binary neutron star injection set using the method described above, with 70 mergers per set. We define a post-merger remnant that can be confidently detected as a system that has an optimal network SNR $>12$~\citep{panther23}.  We compute the number of signals that are accepted into the injection set before the first event with a network SNR $>12$ is identified. The distribution of the number of events added to our sub-threshold sample before an event with SNR $>12$ is found is shown in Fig.~\ref{fig:timetoloud}. We find that in $17/100$ realizations, no SNR $>12$ event is generated before our sample of 70 events is constructed (violet hatched bin at N=70). On the other hand, we find there is a 50\% (57\%) chance that we will observe a single loud post-merger remnant before we detect an ensemble of 25 (35) sub-threshold events.

Our results suggest that time can never mend the substantial chance of detecting a single loud event before the subthreshold population can be used to constrain the maximum neutron star mass using our method. We provide two caveats to this conclusion, though. 

First, while a single, high-amplitude post-merger remnant signal can be used to constrain the nuclear equation of state, we note that represents only a single point estimate. On the other hand, our method allows us to constrain the maximum mass agnostic of the equation of state using a population of mergers. In this sense, we anticipate the measurement of a single remnant signal as well as a population of remnants to provide complementary information, and we envisage a future in which each method inform priors for the other, providing tighter posterior constraints in general.

Second, we note that we expect substantive improvements in the post-merger waveforms models on two related fronts in the coming years that could aid in significant acceleration of our method, in terms of reducing the number of sub-threshold events required to inform the maximum mass. The first front is simply having significantly more post-merger numerical relativity simulations will allow for more robust post-merger waveform approximants; akin to the significant improvement seen in the development of inspiral-merger-ringdown approximants in the last decade. In turn, this may further allow us to use more information from the inspiral portion of the signal (e.g., use the progenitor masses and spins as post-merger priors), which would shrink our post-merger prior volume, and hence provide higher evidence values for signal hypothesis when, in fact, there is a signal present.


\section{Conclusions}\label{sec:conclusions}
Binary neutron star post-merger remnants provide an exciting prospect for measuring the hot nuclear equation of state, complementing nuclear physics experiments as well as astrophysical observations of cold neutron stars. However, there is no comfort in the truth that detection prospects are comparatively grim given 1) the rate at which BNS merger GW detections are made and 2) the relatively low sensitivity of gravitational-wave observatories in the kHz band. In this work, we present a method to combine information about an ensemble of sub-threshold post-merger remnants to learn about the maximum mass of rapidly rotating neutron stars, which can be converted into equation-of-state constraints. We show $\sim25-35$ mergers allows us to constrain the maximum mass to approximately 20\%. 

We are not the first to describe an ensemble method, although we note our proposed technique is different to those of \citet{criswell23} and \citet{Mitra2025}, who both attempt to constrain the neutron star radius assuming quasi-universal neutron star relations between the post-merger peak frequency, the binary chirp mass, and the neutron star radius. Our method is also model dependent, however it suffers from different potential biases. For example, these universal relations can be significantly different if e.g., there are deconfined quarks or other exotic particle species only present at the hot post-merger-remnant temperatures~\citep[e.g.,][]{bauswein19}. We emphasise that our method also suffers from potential biases associated with quasi-universal relations, although this is implicit in the waveform model used to calculate our evidences. In principle, one could ameliorate this potential bias by using agnostic waveforms~\citep[e.g.,][]{clark16,chatziioannou17,easter20}, although this may reduce our potential ability to use the inspiral posterior distributions as priors in future work. 

One of the advantages of our work which combines sub-threshold signals is that it is less sensitive to model misspecification in the waveform modelling. At low signal-to-noise ratios ($\rho < 8$) we are only sensitive to the loudest component of the waveform, $f_\mathrm{peak}$. Second-order features, for example from the tidal disruption of material that forms spiral features, are not as loud as the dominant fundamental mode of the neutron star. However, one can systematically explore the effect of model misspecification on the method by recovering the injected signals with a different waveform model to the one that was used to generate them in the first place. At the singal-to-noise ratio of the injected signals, it is not expected that any substantial systematic bias will be induced on $M_\mathrm{Max}$ provided the waveform model can reliably recover $f_\mathrm{peak}$.

Our work suffers from another potential bias that needs to be considered; the binary neutron star mass distribution. In this work, we assume we know this mass distribution exactly as the $\gtrsim25$ merger events required will provide this information from the ensemble of inspiral parameters. In reality, when performing these measurements one will need to marginalise over the neutron star mass distribution. This will require using combinations of data-driven and parametrically-modeled approaches akin to what is currently happening in the binary black hole literature~\citep[e.g., see][and references therein]{gwtc4pop}. We will therefore not know the mass distribution exactly, implying this will introduce extra uncertainty into our final results.

There is cause to be optimistic about the potential of post-merger gravitational-wave science. Certainly not because of the binary neutron star event rate or the state of federal science funding around the globe, but because the nuclear-physics, numerical-relativity, data-analysis, and instrumentation communities are banding together to understand, detect, and interpret these enigmatic mergers that will hopefully see positive scientific observations in the near future.

\section*{Acknowledgments}
We acknowledge the rightful owners of the land this research was conducted on, the Whadjuk People of the Noongar Nation, and Bunurong Peoples of the Kulin Nation, and pay our respects to elders past and present. We thank the referee for their insightful critique. The authors thank Alexander Criswell for useful comments on the manuscript. FHP is supported by a Forrest Research Foundation Fellowship. This work is supported through Australian Research Council (ARC) Centres of Excellence CE170100004, CE230100016, Discovery Projects DP220101610 and DP230103088, and LIEF Project LE210100002. 
This material is based upon work supported by NSF’s LIGO Laboratory which is a major facility fully funded by the National Science Foundation. Simulations presented in this work was performed on the OzSTAR national facility at Swinburne University of Technology, which receives funding in part from the Astronomy National Collaborative Research Infrastructure Strategy (NCRIS) allocation provided by the Australian Government, and from the Victorian Higher Education State Investment Fund (VHESIF) provided by the Victorian Government.

\section*{Data Availability}
Code and data available on reasonable request to authors.



\bibliographystyle{mnras}
\bibliography{example} 







\bsp	
\label{lastpage}
\end{document}